\documentstyle[preprint,aps,epsf]{revtex}

\def\eqn#1{Eq.~(\ref{#1})}

\def\ipip{ {    {{\bf\cal I}} (m_\pi/p_F^{(p)} ) } }
\def\ipip{ {    {{\bf\cal I}} (m_\pi/p_F^{(p)} ) } }
\def\ikp{{ {{\bf\cal I}} (m_K/p_F^{(p)}) }}
\def\ietap{{ {{\bf\cal I}} (m_\eta/p_F^{(p)}) }}
\def\ipin{{ {{\bf\cal I}} (m_\pi/p_F^{(n)}) }}
\def\ikn{{  {{\bf\cal I}} (m_K/p_F^{(n)}) }}
\def\ietan{{  {{\bf\cal I}} (m_\eta/p_F^{(n)}) }}
\def\denp{{ d^{(p)} }}
\def\denn{{ d^{(n)} }}
\def\pfn{{ p_F^{(n)}} }
\def\pfp{{ p_F^{(p)} }}
\def\dpip{{ D ( p_F^{(p)}/m_\pi ) }}
\def\dpin{{ D ( p_F^{(n)}/m_\pi ) }}
\def\dkp{ {D ( p_F^{(p)}/m_K ) }}
\def\dkn{ {D ( p_F^{(n)}/m_K ) }}
\def\detap{{ D ( p_F^{(p)}/m_\eta ) }}
\def\detan{ {D ( p_F^{(n)}/m_\eta ) }}

\begin{document}

{\tighten
\preprint{\vbox{\hbox{CALT-68-2004}
\hbox{CMU-HEP95-09}
\hbox{DOE-ER/40682-99}
\hbox{DOE Research and }
\hbox{Development Report}
}}

\title{Hyperon Masses in Nuclear Matter \footnote{Work supported
in part by
the U.S. Dept. of Energy under Grant No. DE-FG03-92-ER40701 and
DE-FG02-91-ER40682.}}

\author{Martin J. Savage}
\address{Department of Physics, Carnegie Mellon University\\
Pittsburgh, Pennsylvania 15213
U.S.A.\\ {\tt savage@thepub.phys.cmu.edu}}
\author{Mark B. Wise}
\address{California Institute of Technology\\
Pasadena, CA 91125
U.S.A.\\ {\tt wise@theory.caltech.edu}}

\bigskip
\date{July 1995}

\maketitle
\begin{abstract}

We analyze hyperon and nucleon mass shifts in nuclear matter
using chiral
perturbation theory.
Expressions for the mass shifts that include strong
interaction effects at
leading order in the density are derived.
Corrections to our results are suppressed by powers of the Fermi
momentum
divided by either the chiral symmetry breaking scale or the
nucleon mass.  Our
work is relevant for neutron stars and for large hypernuclei.
\end{abstract}
\vfill\eject

In nuclear matter the masses of hyperons and nucleons are shifted
from their
vacuum values by strong interaction effects.
Understanding these mass shifts
is important for neutron stars
\cite{g85,EKO91,PPLP92,PCL95,PBPELK95,KPE95,PT95}
and for large hypernuclei (see for example \cite{D92,GD95}).
Most approaches to this problem rely on the use of
phenomenological
models
\cite{g85,EKO91,PCL95,PBPELK95,KPE95,GD95,CW91,MJa94,MJ95,JF94,JF
N95,JN95}
since the relevant strong interaction effects involve QCD in the
nonperturbative
regime.
In this letter we will consider mass shifts
for the baryon octet using a model independent approach to
baryon-baryon interactions based on chiral perturbation theory.
The price to be paid for model independence is
that an expansion in momentum is necessary and that
our expressions for some of the mass shifts involve parameters in
the heavy baryon chiral Lagrangian that have not yet been
determined from
experiment.
However, there are mass relations that hold independent of these
parameters
in analogy with the Gell-Mann--Okubo mass formula for SU(3)
breaking.

We study changes in the masses of the baryon octet
$(n,p,\Lambda,\Sigma, \Xi)$
that arise from their strong interactions with a degenerate Fermi
gas of
neutrons at density $d^{(n)} = (p_F^{(n)})^3/3\pi^2$ and protons
at density
$d^{(p)} = (p_F^{(p)})^3/3\pi^2$ (i.e., nuclear matter).
These interactions
are determined by a heavy-baryon chiral Lagrangian that is
consistent with the
$SU(3)_L \times SU(3)_R$ chiral symmetry of QCD.  An expansion in
derivatives
is
appropriate for Fermi momenta $p_F^{(n)}$ and $p_F^{(p)}$ that
are small
compared with the chiral symmetry breaking scale and the baryon
masses.

In the rest frame of background nuclear matter the energy of a
baryon and its
three-momentum are related (for $|\vec p| << M_B$)  by
$E^{\rm NM} = M_B + \Delta \tilde M_B + |\vec p|^2/2M_B + \eta_B
|\vec p|^2 +
...$
which is to be compared to the free space relation
$E^{\rm vac} = M_B  + |\vec p|^2/2M_B + ...$
where the dots denote terms higher order in the momentum
expansion.
We can combine the effects of the medium with the free space
mass, $M_B$,
by defining
$M^*_B$ and $\tilde M_B$, with $\tilde M_B = M_B+\Delta \tilde
M_B$ and
$M^*_B = M_B ( 1+ 2\eta_B M_B)^{-1} = M_B + \Delta M_B^*$
allowing us to write
$E^{\rm NM} = \tilde M_B  + |\vec p|^2/2M_B^* + ... .$
The quantities $\Delta \tilde M_B$ and $\eta_B$ arise from
interactions with
the
background medium
and in general $M_B^*\ne \tilde M_B.$
We compute the lowest order contribution to the momentum
independent
mass shift $\Delta\tilde M_B$ and the coefficient of the $ |\vec
p|^2 $ term
$\eta_B$ for the octet baryons using chiral perturbation theory.

The pseudo-Goldstone boson fields $(\pi, K, \eta)$ can be written
as a
$3\times 3$ special unitary matrix
\begin{equation}
\Sigma = \exp {2i{\bf \Pi}\over f} \ \ \ ,
\end{equation}
where
\begin{equation}\label{mesonfield}
{\bf\Pi} =   \left[\matrix{\pi^0/ \sqrt{2} + \eta/\sqrt{6} &
\pi^+
& K^+ \cr
\pi^- & -\pi^0/\sqrt{2} + \eta/\sqrt{6} & K^0\cr
K^- & \bar K^0 & -\sqrt{2\over 3}\eta\cr}\right] \ \ \ .
\end{equation}
Under chiral $SU(3)_L \times SU(3)_R$ symmetry, $\Sigma
\rightarrow L \Sigma
R^\dagger$ where $L \epsilon SU(3)_L$ and $R\epsilon SU (3)_R$.
At leading
order in chiral perturbation theory $f$ can be identified with
the pion decay
constant $(f_\pi \simeq 132 MeV)$. When describing the
interactions of the pseudo-Goldstone bosons with other fields it
is convenient
to introduce
\begin{equation}
\xi = \exp {i{\bf\Pi}\over f} = \sqrt{\Sigma} \ \ \ .
\end{equation}
Under chiral $SU(3)_L \times SU(3)_R$ symmetry
\begin{equation}
\xi \rightarrow L \xi U^\dagger = U \xi R^\dagger \ \ \ ,
\end{equation}
where in general $U$ is a complicated function of $L,R$ and the
meson fields
${\bf \Pi}$.  For transformations $V = L = R$ in the unbroken
$SU(3)_V$
subgroup $U = V$.

The baryon fields are introduced as a $3\times 3$ octet matrix
\begin{equation}\label{baryons}
B = \left[\matrix{\Sigma^0/\sqrt{2} + \Lambda/\sqrt{6} &
\Sigma^+ & p\cr
\Sigma^- & -\Sigma^0/\sqrt{2} + \Lambda/\sqrt{6} & n\cr
\Xi^- & \Xi^0 & -\sqrt{{2\over 3}} \Lambda \cr}\right]\ \ \ ,
\end{equation}
that transforms under chiral $SU(3)_L \times SU(3)_R$ as
\begin{equation}\label{baryontrans}
B \rightarrow UBU^\dagger \ \ \ .
\end{equation}

We construct a chiral Lagrangian by treating the baryons as heavy
static
2-component fields.  The chiral Lagrangian is written as
\begin{equation}
{\cal L} = {\cal L}^{(1)} + {\cal L}^{(2)} + \dots\ \ \ ,
\end{equation}
where ${\cal L}^{(j)}$ contains $2j$ baryon fields.

${\cal L}^{(1)}$ is the familiar heavy baryon chiral Lagrangian
that gives the
interactions of the baryon octet with the pseudo-Goldstone
bosons.  At leading
order in chiral perturbation theory
\begin{eqnarray}\label{NRlag}
{\cal L}^{(1)} & = &  Tr B_j^\dagger i\partial_0 B_j + i Tr
B_j^\dagger [V_0,
B_j]\ \nonumber\\
&  & - D Tr B_j^\dagger \vec \sigma_{jk} \{\vec A, B_k\} - F Tr
B_j^\dagger
\vec
\sigma_{jk} [\vec A, B_k] \ \ \ ,
\end{eqnarray}
with the repeated spin indices $j$ and $k$ summed over 1,2 and
the vector and
axial-vector chiral fields are
\begin{eqnarray}\label{VAfields}
V_\mu & = &  {1\over 2} (\xi^\dagger \partial_\mu \xi + \xi
\partial_\mu
\xi^\dagger)\ \ ,\\
A_\mu &  = &  {i\over 2} (\xi^\dagger \partial_\mu \xi - \xi
\partial_\mu
\xi^\dagger)\ \ \ .
\end{eqnarray}
Nuclear $\beta$ decay and semileptonic hyperon decay give $F
\simeq 0.44$ and
$D \simeq 0.81$ at tree-level.
At higher order in chiral perturbation theory there are
corrections to \eqn{NRlag}\   coming from terms with more
derivatives and terms
with
insertions of the light quark mass matrix.

Interactions between baryons mediated by pseudo-Goldstone boson
exchange give
rise to
hyperon mass shifts.
At the same order in chiral perturbation
theory (i.e., order $p_F^3$) terms in ${\cal L}$ with four baryon
fields (and
no derivatives) also play a role.  They are given by
\begin{eqnarray}\label{fourb}
{\cal L}^{(2)} =
&-&{c_1\over f^2} Tr (B_i^\dagger B_i B_j^\dagger B_j)
- {c_2\over f^2} Tr (B_i^\dagger B_j B_j^\dagger B_i)\nonumber\\
&-& {c_3\over f^2} Tr (B_i^\dagger B_j^\dagger B_i B_j)
- {c_4\over f^2} Tr (B_i^\dagger B_j^\dagger B_j B_i)\nonumber\\
&-& {c_5\over f^2} Tr (B_i^\dagger B_i) Tr (B_j^\dagger B_j)
- {c_6\over f^2} Tr (B_i^\dagger B_j) Tr (B_j^\dagger B_i) \ \ \
{}.
\end{eqnarray}
Factors of $1/f^2$ appear in \eqn{fourb}\  so that the
coefficients $c_j$ are
dimensionless.
At higher order in chiral perturbation theory there are
corrections to \eqn{fourb}\  coming from terms with derivatives
and terms with
insertions of the light quark mass matrix.
There are also contributions to the baryon mass shifts
from operators in ${\cal L}^{(j)}$, $j > 2$, involving more
baryon fields,
e.g. $Tr( B^\dagger_i B_i B^\dagger_j B_j B^\dagger_k B_k ) $,
but these are suppressed by additional powers of
$d^{(n,p)}/(4\pi f^3) $ over the contribution from
operators appearing in \eqn{fourb}\ .
Note there is no term involving
$Tr (B_i^\dagger B_j^\dagger) \ Tr ( B_i B_j ) $ in \eqn{fourb}\
as it can be eliminated using a Cayley-Hamilton operator
identity:
\begin{eqnarray}
& & - Tr (B_i^\dagger B_j^\dagger B_i B_j)
+ Tr(B_i^\dagger B_j^\dagger B_j B_i)
-  {1\over 2} Tr (B_i^\dagger B_j B_j^\dagger B_i)
+ {1\over 2} Tr (B_i^\dagger B_i B_j^\dagger B_j)\nonumber\\
& &= {1\over 2} Tr (B_i^\dagger B_i) Tr (B_j^\dagger B_j)
- {1\over 2} Tr (B_i^\dagger B_j) Tr (B_j^\dagger B_i)
- {1\over 2} Tr (B_i^\dagger B_j^\dagger) Tr (B_i B_j)\ \ \ ,
\end{eqnarray}
It is easy to understand from spin-flavour considerations
why there are only six independent four-baryon operators.
The  $SU(3)_V$ decompostion of the product of two baryon octets
is
${\bf 8}\otimes {\bf 8} = {\bf 27}\oplus{\bf 10
}\oplus{\bf\overline{10}}
\oplus{\bf 8}_A
\oplus{\bf 8}_S\oplus{\bf 1}$ of which the
${\bf 27} , {\bf 8}_S , {\bf 1}$ are symmetric and the
${\bf 10 } , {\bf\overline{10}} , {\bf 8}_A$ are antisymmetric
under
baryon interchange.
Fermi statistics require that the baryons in the
${\bf 27} , {\bf 8}_S , {\bf 1}$ have spin $S=0$ and those in the
${\bf 10 } , {\bf\overline{10}} , {\bf 8}_A$ have $S=1$ as the
operators under
consideration give rise to S-wave interactions only.
Hence, there are only 6 independent four-baryon operators that
exist at
this order in chiral perturbation theory.
It is these six operators (and pseudo-Goldstone boson exchange)
that give the leading contribution to S-wave
baryon-baryon scattering, such as $\Lambda N\rightarrow \Lambda
N$,
in chiral perturbation theory.  An $SU(3)_V$ analysis of
hyperon-nucleon
scattering can be
found in \cite{DF90}.

The four-baryon operators in \eqn{fourb}\ include effects from
the
repulsive core of the baryon-baryon potential.
For example, a repulsive spherical well of infinite height and
width
$r_0$ gives rise to coefficients, $c_j$, of order $4\pi f^2
r_0/M_B$.

Using \eqn{NRlag}\  and \eqn{fourb}\  we find the following one
loop
expressions
for the
baryon octet mass shifts in nuclear matter:
\begin{eqnarray}\label{delma}
\Delta\tilde M_n &=&  {d^{(n)}\over f^2} \Bigg\{  {(F+D)^2\over
4} {\bf{\cal
I}}
(m_\pi/p_F^{(n)}) + {(3F-D)^2\over 12} {{\bf\cal I}}
(m_\eta/p_F^{(n)})+ c_1 -
c_2 + c_5 - c_6 \Bigg\}\nonumber\\
& &+ {d^{(p)}\over f^2} \Bigg\{  {(F+D)^2\over 2} {{\bf\cal I}}
(m_\pi/p_F^{(p)}) + 2c_1 + c_2 + 2c_5 + c_6 \Bigg\} \nonumber\\
\Delta \tilde M_p &=& {d^{(n)}\over f^2} \Bigg\{  {(F+D)^2\over
2} {{\bf\cal
I}}
(m_\pi/p_F^{(n)}) + 2c_1 + c_2 + 2c_5 + c_6 \Bigg\}\nonumber\\
& & + {d^{(p)}\over f^2} \Bigg\{ {(F + D)^2\over 4} {{\bf\cal I}}
(m_\pi/p_F^{(p)}) + {(3F - D)^2\over 12} {{\bf\cal I}} (m_\eta
/p_F^{(p)}) +
c_1 - c_2 + c_5 - c_6 \Bigg\} \nonumber\\
\Delta \tilde M_{\Sigma^{+}} &=& {d^{(n)}\over f^2} \left\{
{c_3\over 2} + c_4
+ 2c_5
+ c_6\right\}\nonumber\\
& &+ {d^{(p)}\over f^2} \left\{ {(D - F)^2\over 2} {{\bf\cal I}}
(m_K/p_F^{(p)}) - c_1 - 2c_2 + 2c_5 + c_6\right\} \nonumber\\
\Delta\tilde M_{\Sigma^{-}} &=& {d^{(n)}\over f^2} \left\{
{(D-F)^2\over 2}
{{\bf\cal I}} (m_K/p_F^{(n)}) -c_1 -2c_2  + 2c_5 + c_6\right\}
\nonumber\\
& & + {d^{(p)}\over f^2} \left\{ {c_3\over 2} + c_4 + 2c_5 + c_6
\right\}
\nonumber\\
\Delta \tilde M_{\Xi^{0}} &=&  {d^{(n)}\over f^2} \left\{ c_3 +
2c_4 + 2c_5 +
c_6
\right\} + {d^{(p)}\over f^2} \left\{ {1\over 2} c_3 + c_4 + 2c_5
+ c_6\right\}
\nonumber\\
\Delta \tilde M_{\Xi^{-}} &=& {d^{(n)}\over f^2} \left\{ {1\over
2} c_3 + c_4 +
2c_5 + c_6 \right\} + {d^{(p)}\over f^2} \left\{ c_3 + 2c_4 +
2c_5 + c_6
\right\}
\ \ \ \ .
\end{eqnarray}
For the $\Lambda$ and $\Sigma^0$ there is a $2 \times 2$ mass
matrix with
entries
\begin{eqnarray}\label{delmb}
\Delta \tilde M_{\Lambda\Lambda} &=& {d^{(n)}\over f^2} \Bigg\{
{1\over 12}
(D+
3F)^2 {{\bf\cal I}} (m_K/p_F^{(n)}) + {7\over 6} c_1 + {1\over 3}
c_2 +
{13\over 12} c_3 + {7\over 6} c_4 + 2c_5 + c_6 \Bigg\}\nonumber\\
& & + {d^{(p)} \over f^2}
\Bigg\{ {1\over 12} (D + 3F)^2 {{\bf\cal I}} (m_K/p_F^{(p)}) +
{7\over 6} c_1
+ {1\over 3} c_2 + {13\over 12} c_3 + {7\over
6} c_4 + 2c_5 + c_6 \Bigg\}\nonumber \\
\Delta \tilde M_{\Sigma^{0} \Sigma^{0}} &=& {d^{(n)}\over f^2}
\Bigg\{ {(D -
F)^2\over 4} {{\bf\cal I}} (m_K/p_F^{(n)}) - {1\over 2} c_1 - c_2
+ {1\over 4}
c_3 + {1\over 2} c_4 + 2c_5 + c_6\Bigg\}\nonumber\\
& &+ {d^{(p)}\over f^2} \Bigg\{ {(D - F)^2\over 4} {{\bf\cal I}}
(m_K/p_F^{(p)}) - {1\over 2} c_1 - c_2 + {1\over 4} c_3 + {1\over
2} c_4 + 2c_5
+ c_6\Bigg\} \nonumber\\
\Delta \tilde M_{\Lambda\Sigma} &=& \Delta M_{\Sigma\Lambda} =
{d^{(n)}\over
f^2}
\Bigg\{{(D + 3F) (D-F)\over 4\sqrt{3}} {{\bf\cal I}}
(m_K/p_F^{(n)}) +
{c_1\over 2\sqrt{3}} + {c_2 \over \sqrt{3}} - {5\over 4\sqrt{3}}
c_3 - {1\over
\sqrt{3}} c_4 \Bigg\} \nonumber\\
& &+ {d^{(p)}\over f^2} \Bigg\{ {(D + 3F) (F-D)\over 4\sqrt{3}}
{{\bf\cal I}}
(m_K/p_F^{(p)}) - {c_1\over 2\sqrt{3}} - {c_2\over \sqrt{3}} +
{5\over
4\sqrt{3}} c_3 + {1\over \sqrt{3}} c_4 \Bigg\} \ \ \ .
\end{eqnarray}
In \eqn{delma} and \eqn{delmb} the function ${{\bf\cal I}} (x)$
is defined by
\begin{equation}
{{\bf\cal I}}(x) = 1 - 3x^2 + 3x^3 Arctan (1/x)\ \ \ .
\end{equation}

Naively, corrections to \eqn{delma} and \eqn{delmb} are
suppressed by powers of
the
Fermi momentum divided
by the chiral symmetry breaking scale (or the nucleon mass) and
by powers of
the light quark masses divided by the chiral symmetry breaking
scale (or the
nucleon mass).  This is essentially a consequence of dimensional
analysis as
Feynman diagrams with more than one  loop constructed from the
vertices of the
Lagrange densities in \eqn{NRlag}\  and \eqn{fourb}\  contain
more powers of
$1/f$ which must
be compensated by additional factors of  $p_F$ or light quark
masses.
Corrections to the Lagrange densities in \eqn{NRlag}\  and
\eqn{fourb}\
from operators with
more derivatives (or insertions of the light quark mass matrix)
have
coefficients with more powers of $1/f$ which again must be
compensated by
additional factors of $p_F$ or light quark masses.
However, this power
counting is not quite correct.  Feynman diagrams with more baryon
propagators
than loops have infrared divergent loop integrations over energy.
 Furthermore,
insertions of the kinetic energy, ${\cal L}_{KE} = Tr B_i^\dagger
\nabla^2
B_i/2M_B$, increase the number of baryon propagators but not the
number of
loops (this correction to the Lagrange density in \eqn{NRlag}\
cannot be
treated as
perturbation).  The cure for these problems is to sum insertions
of the kinetic
energy changing the baryon propagator to
\begin{equation}\label{propmom}
i \left[ {\Theta (p_F - |\vec p|)\over p^0 - \vec p^2/2M_B -
i\epsilon} +
{\Theta (|\vec p | - p_F)\over p^0 - \vec p ^2/2M_B + i\epsilon}
\right]\ \ \ .
\end{equation}
Then infrared divergent $p^0$ integrations are cut off by the
kinetic energy
which produces factors of $M_B$ in the numerator.  At $l$-loops
one finds
corrections suppressed by a factor of order $(M_B p_F/(4\pi
f^2))^{\ell - 1}$.
As a consequence,
the leading correction to our results arise from two loop
diagrams and are
suppressed by only $\sim M_B p_F/(4\pi f^2)$ (instead of $\sim
p_F^2/(4\pi
f)^2$).
At nuclear density ($p_F\sim 350 $ MeV/c) this factor is not
small.
Therefore the leading order results derived here may receive
significant
corrections
from higher orders in chiral perturbation theory.

We adopt power counting rules where $M_B$ is considered to be of
the same order
as the chiral symmetry breaking scale $\sim 4 \pi f$ and two
factors of $p_F$
are considered to be of the same order as a light quark mass
(Recall that the
squares of the pseudo-Goldstone boson masses are proportional to
light quark
masses).  Consequently in \eqn{delma}\ and \eqn{delmb}\  the full
dependence of
the
baryon mass shifts on $(p_F/\mu)$ where $\mu = m_K, m_\pi$ or
$m_\eta$ is kept.

The baryon mass shifts given in \eqn{delma}\ and \eqn{delmb}\ are
our
principal results.
If the nuclear matter is not
isospin symmetric (i.e., $d^{(n)} \not= d^{(p)})$ there is
$\Lambda -
\Sigma^{0}$ mixing (\eqn{delmb}).  More importantly, the eight
octet masses
are expressed in terms of six coefficients $c_j$ so there are
mass relations
independent of the
coefficients of the four baryon operators.  We find that
\begin{equation}
2\Delta\tilde M_{\Sigma^{0} \Sigma^{0}} - \Delta \tilde
M_{\Sigma^{+}} - \Delta
\tilde M_{\Sigma^{-}} = 0 \ \ \ ,
\end{equation}
which is independent of the composition of the nuclear matter, it
is true
for an arbitrary proton to neutron ratio.
A second mass relation is given in terms of the mesonic
contribution
\begin{equation}
d^{(n)} \left(\Delta \tilde M_{\Sigma^{+}} - \Delta \tilde
M_{\Xi^{-}}\right) +
d^{(p)}
\left(\Delta
\tilde M_{\Xi^{0}} - \Delta \tilde M_{\Sigma^{-}} \right) =
{(D-F)^2\over 2
f^2} \denn\denp
\left( \ikp-\ikn \right)  \ \ \ .
\end{equation}
We see that in the extreme case of neutron matter ($\denp = 0$)
this relation becomes
$\Delta \tilde M _{\Sigma^{+}}~=~\Delta \tilde M_{\Xi^{-}}$.
There is also a relation involving the $\Sigma^0\Lambda$ mixing
term
\begin{eqnarray}
f^2 & d^{(n)}  & \left( -3\Delta \tilde M_{\Lambda\Lambda}
+ 2 \Delta \tilde M_{p} - 2\sqrt{3} \Delta \tilde
M_{\Lambda\Sigma}
-2 \Delta \tilde M_{\Sigma^-} + 3 \Delta \tilde
M_{\Sigma^0\Sigma^0}
\right)\nonumber\\
+\  f^2 & d^{(p)} & \left(  3\Delta \tilde M_{\Lambda\Lambda} -
2\Delta \tilde
M_{n}
- 2\sqrt{3}\Delta \tilde M_{\Lambda\Sigma}
-2 \Delta \tilde M_{\Sigma^-}+ \Delta \tilde
M_{\Sigma^0\Sigma^0}\right)
\nonumber\\
& = &  (D+F)^2 \left[ \denn^2+\denp^2-{1\over 2}\denn\denp\right]
\left(\ipin-\ipip\right)\nonumber\\
& &  - (D+F)(D-3F)\ \denn\denp \left( \ikn-\ikp \right)
\nonumber\\
& &  - (D+F)^2 \left( \denn^2 \ikn - \denp^2 \ikp
\right)\nonumber\\
& &  - {1\over 6} (D-3F)^2\  \denn\denp \left( \ietan-\ietap
\right)
\end{eqnarray}
When $d^{(p)} = d^{(n)}$ the background nuclear matter is isospin
symmetric
leading to  baryons in the same isomultiplet receiving the same
mass shift
and $\Sigma^0 \Lambda$ mixing vanishing.

Weinberg \cite{W90,W91} has extracted values for $c_2 + c_6  =
f^2 C_T$ and
$2c_1 + c_2 + 2 c_5 + c_6 =  f^2 C_S$ from the measured S-wave
nucleon
scattering lengths .  He found $C_T \approx (1/45 {\rm MeV})^2$
and $C_S
\approx -
(1/88 {\rm MeV})^2$ which  give very large nucleon mass shifts at
nuclear
density \cite{MPW92}.
If \eqn{delma}\ and \eqn{delmb}\ are to be relevant at nuclear
density
the coefficients $C_{S,T}$
must be smaller than these values.
It may be possible to extract other combinations of the
coefficients $c_i$ by
measuring the S-wave scattering lengths for hyperon nucleon
scattering.
For hyperon-proton strong interactions the relevant spin singlet
$a_{S=0}$
and spin triplet $a_{S=1}$ elastic scattering lengths are
\begin{eqnarray}
| a_{S=0} (\Lambda p\rightarrow\Lambda p)| &  & = {1\over 2\pi
f^2}
{M_p M_\Lambda\over M_p+M_\Lambda }
|{5\over 3} c_1 - {5\over 3} c_2 - {1\over 6} c_3
+ {1\over 6}c_4  + 2 c_5 - 2 c_6|\nonumber\\
|a_{S=1}(\Lambda p\rightarrow\Lambda p)| & &
= {1\over 2\pi f^2} {M_p M_\Lambda\over M_p+M_\Lambda }
| c_1 + c_2 +{3\over 2} c_3 + {3\over 2} c_4  + 2 c_5 + 2
c_6|\nonumber\\
|a_{S=0}(\Sigma^+ p\rightarrow\Sigma^+ p)| & &
= {1\over 2\pi f^2} {M_p M_\Sigma\over M_p+M_\Sigma}
2|c_1-c_2+c_5-c_6|\nonumber\\
|a_{S=1}(\Sigma^+ p\rightarrow\Sigma^+ p)| & &
= {1\over 2\pi f^2} {M_p M_\Sigma\over M_p+M_\Sigma }
2|-c_1-c_2+c_5+c_6|\nonumber\\
|a_{S=0}(\Sigma^- p\rightarrow\Sigma^- p)| & &
= {1\over 2\pi f^2} {M_p M_\Sigma\over M_p+M_\Sigma }
|-c_3 + c_4 + 2 c_5 - 2 c_6|\nonumber\\
|a_{S=1}(\Sigma^- p\rightarrow\Sigma^- p)| & &
= {1\over 2\pi f^2} {M_p M_\Sigma\over M_p+M_\Sigma }
|c_3 + c_4 + 2 c_5 + 2 c_6|\nonumber\\
|a_{S=0}(\Xi^- p\rightarrow\Xi^- p)| & &
= {1\over 2\pi f^2} {M_p M_\Xi\over M_p+M_\Xi }
2|-c_3 + c_4 + c_5 - c_6|\nonumber\\
|a_{S=1}(\Xi^- p\rightarrow\Xi^- p)| & &
= {1\over 2\pi f^2} {M_p M_\Xi\over M_p+M_\Xi }
2|c_3 + c_4 + c_5 + c_6|\nonumber\\
|a_{S=0}(\Xi^0 p\rightarrow\Xi^0 p)| & &
= {1\over 2\pi f^2} {M_p M_\Xi\over M_p+M_\Xi }
|c_3 - c_4 -2c_5+2c_6|\nonumber\\
|a_{S=1}(\Xi^0 p\rightarrow\Xi^0 p)| & &
= {1\over 2\pi f^2} {M_p M_\Xi\over M_p+M_\Xi }
|c_3 + c_4 + 2c_5 + 2c_6|
\end{eqnarray}
Because of the derivative coupling in \eqn{NRlag}\ tree-level
pseudo-Goldstone boson exchange
doesn't contribute to the scattering lengths.

The baryon mass shifts given in \eqn{delma}\ and \eqn{delmb}\
may be of use for large
hypernuclei and for neutron stars.
Assuming a large nucleus can be modeled by a degenerate Fermi
gas replacing one
of its neutrons with a $\Lambda$  results in a hypernucleus
with a mass shift that
is partly determined by $\Delta \tilde M_\Lambda - \Delta \tilde
M_n$.
Study of the  S-wave hyperon-nucleon scattering lengths
may eventually lead to a determination of the coefficients
$c_j$ which can then be used to predict hyperon mass shifts
in neutron stars
(where $d^{(p)} \ll d^{(n)}$) and in large hypernuclei.

The mass that appears in the kinetic energy
$M_B^*$ differs from the vacuum mass $M_B$ through momentum
dependent
interactions with the background nuclear matter that give a
non-zero value for
the coefficients $\eta_B$.
At one loop the four-baryon operators in \eqn{fourb}\ do not
contribute to
$\eta_B$
as they are independent of the baryon momentum.
However, the meson graphs do give a calculable contribution to
$\eta_B$
and we find that
\begin{eqnarray}\label{dyneta}
\eta_n & = &  {\pfn\over 6 \pi^2 f^2 } \left( {1\over 2} ( D+F)^2
\dpin
+ {1\over 6} (D-3F)^2 \detan \right) \nonumber\\
& & + {\pfp\over 6 \pi^2 f^2 } (D+F)^2 \dpip \nonumber \\
\eta_p & = &   {\pfn\over 6 \pi^2 f^2 } (D+F)^2 \dpin
\nonumber\\
& &+ {\pfp\over 6 \pi^2 f^2 } \left( {1\over 2} ( D+F)^2 \dpip
+ {1\over 6} (D-3F)^2 \detap \right) \nonumber \\
\eta_{\Lambda\Lambda} & = &  {1\over 6 \pi^2 f^2 } {1\over 6} (
D+3F)^2
\left( \pfn \dkn  + \pfp\dkp \right) \nonumber \\
\eta_{\Sigma^+} & = &  {\pfp\over 6 \pi^2 f^2 } (D-F)^2 \dkp
\nonumber\\
\eta_{\Sigma^0\Sigma^0} & = &  {1\over 6 \pi^2 f^2 }  {1\over 2}
( D-F)^2
\left( \pfn\dkn  + \pfp\dkp \right) \nonumber \\
\eta_{\Sigma^-} & = &  {\pfn\over 6 \pi^2 f^2 } ( D-F)^2 \dkn
\nonumber \\
\eta_{\Sigma\Lambda} = \eta_{\Lambda\Sigma}
& = &  {1\over 6 \pi^2 f^2 } {( D-F)(D+3F)\over2\sqrt{3}}
\left( \pfn\dkn - \pfp\dkn \right)
\end{eqnarray}
where the function $D(x)$ is
\begin{eqnarray}\label{dynfun}
D(x) = {x^2\over (1+x^2)^2 }\ \ \ \ .
\end{eqnarray}
For  proton-neutron symmetric nuclear matter with
Fermi levels $\pfn = \pfp = 270 {\rm MeV/c}$ we find that
\begin{eqnarray}
\Delta M_n^* & =& \Delta M_p^* = -154 {\rm MeV}\ \ ,\ \
\Delta M_\Lambda^* = -152 {\rm MeV} \nonumber\\
\Delta M_{\Sigma^+}^* & = & \Delta M_{\Sigma^0}^*
= \Delta M_{\Sigma^-}^* = -18 {\rm MeV} \nonumber\\
\Delta M_{\Xi^0}^* & = & \Delta M_{\Xi^-}^* = 0
\end{eqnarray}
and there is no $\Sigma^0-\Lambda$ mixing.
The values for $\Delta M_n^*$ and $\Delta M_p^*$
are in qualitative agreement with those extracted from nuclei
(see, e.g. \cite{BM69}).
For neutron matter with $\pfn = 350 {\rm MeV/c}$ we find that the
mass shifts are
\begin{eqnarray}
\Delta M_n^* & = & -57 {\rm MeV}\ \ ,\ \ \Delta M_p^* = -100 {\rm
MeV}
\nonumber\\
\Delta M_\Lambda^* & = & -126 {\rm MeV}\ \ ,\ \
\Delta M_{\Sigma^+}^* = 0 {\rm MeV}\nonumber\\
\Delta M_{\Sigma^0}^*  & = & -15 {\rm MeV}\ \ ,\ \
\Delta M_{\Sigma^-}^* = -30 {\rm MeV} \nonumber\\
\Delta M_{\Xi^0}^* & = & \Delta M_{\Xi^-}^* = 0\ \ \ .
\end{eqnarray}

The utility of our results depends crucially upon the
smallness of higher orders
in chiral perturbation theory at nuclear density.
A computation at next order in chiral perturbation theory
may help resolve this issue.
We hope to present results on this in a future publication.

\bigskip

\centerline{\bf Acknowledgements}

MJS would like to thank David Kaplan and Tom Cohen for
enlightening conversations.
We would also like to thank M. Prakash and C. Pethick for
comments on the manuscript.
MJS would like to thank the High Energy Physics Group at
Caltech and the
Institute for Nuclear theory at the University of Washington
for kind hospitality during the course of this work.


\begin{references}

\bibitem{g85} N.K. Glendenning, Ap. J. {\bf 293}, 470 (1985);
Nuc. Phys. {\bf A493}, 521 (1989).

\bibitem{EKO91}J. Ellis, J.I. Kapusta and K.A. Olive,
Nucl. Phys. {\bf B348}, 345 (1991).

\bibitem{PPLP92}M. Prakash et al., Ap. J. {\bf 390}, L77 (1992).

\bibitem{PCL95} M. Prakash, J. Cooke and J.M. Lattimer,
Phys. Rev. {\bf D52}, 661 (1995).

\bibitem{PBPELK95} M. Prakash et al., Phys. Rep. (1995) ,
to be published.

\bibitem{KPE95} R. Knorren, M. Prakash and P.J. Ellis,
SUNY-NTG-95-15, (1995).

\bibitem{PT95} C.J. Pethick and V. Thorsson,
{\it The Lives of the Neutron Stars},
edited by M.A. Alpar et al.,p121, (1995).

\bibitem{D92} C.B. Dover, Nucl. Phys. {\bf A547} , 27c (1992).

\bibitem{GD95} A. Gal and C.B. Dover, Nucl. Phys.
{\bf A585}, 1c (1995),
and references therein.

\bibitem{CW91} J. Cohen and H.J. Weber, Phys. Rev.
{\bf C44}, 1181 (1991).

\bibitem{MJa94} J. Mares and B.K. Jennings, Phys. Rev.
{\bf C49}, 2472 (1994).

\bibitem{MJ95} J. Mares and B.K. Jennings, Nucl. Phys.
{\bf A585}, 347c (1995).

\bibitem{JF94} X. Jin and R.J. Furnstahl, Phys. Rev.
{\bf C49}, 1190 (1994).

\bibitem{JFN95} X. Jin, R.J. Furnstahl and M. Nielson,
Nucl. Phys. {\bf A585}, 333c (1995).

\bibitem{JN95} X. Jin and M. Nielsen, Phys. Rev. {\bf C51},
347 (1995).

\bibitem{DF90} C.B. Dover and H. Feshbach, Ann. of Phys.
{\bf 198}, 321 (1990).

\bibitem{W90} S. Weinberg, Phys. Lett. {\bf B251}, 288 (1990).

\bibitem{W91} S. Weinberg, Nucl. Phys. {\bf B363}, 3 (1991).

\bibitem{MPW92} D. Montano, H.D. Politzer and M.B. Wise,
Nucl. Phys. {\bf B375}, 507 (1992).

\bibitem{BM69} A. Bohr and B.R. Mottelson, {\it Nuclear
Structure},
W.A. Benjamin Inc. Publishing, 1969.
\end{references}
\end{document}